# Energy Aware Task Scheduling for Soft Real Time Systems using an Analytical Approach for Energy Estimation

Namita Sharma[1], Vineet Sahula[2], and C.P. Ravikumar[3]

[1,2] Malaviya National Institute of Technology, Jaipur, INDIA.

[3] Texas Instruments, Bangalore, INDIA.

[1] namita@cse.iitd.ac.in,  [2] sahula@ieee.org,  [3] rkumar@ti.com

**ABSTRACT**

Embedded systems have pervaded all walks of our life. With the increasing importance of mobile embedded systems and flexible applications, considerable progress in research has been made for power management. Power constraints are increasingly becoming the critical component of the design specifications of these systems. It helps in pre-determining the suitable hardware architecture for the target application. Very Large Instruction Word (VLIW) processors provide a means to efficiently exploit the instruction level parallelism (ILP) exhibited by a significant segment of embedded applications. Circuit level or gate level power analysis techniques prove to be impractical for the power cost estimation of the software component of the system. The aim of this paper is to present a technique to estimate 'pre-run time' and 'power' of a software mapped onto a hardware system; guaranteeing the compliance of temporal constraints while generating a schedule of tasks of software. Real time systems must handle several independent macro-tasks, each represented by a task graph, which includes communications and precedence constraints. We propose a novel approach for power estimation of embedded software using the Control Data Flow Graph (CDFG) or task graph model. This methodology uses an existing Hierarchical Concurrent Flow Graph (HCFG) technique for the power analysis of the CDFGs. We have evaluated our technique for energy efficient scheduling over various task graph benchmarks using Trimaran, an environment for software characterization and PrimePower from Synopsys has been used to obtain power estimates for the elementary functional units of datapath. The results obtained prove the utility and efficacy of our proposed approach for power analysis of embedded software. We present a methodology to obtain an energy optimal voltage assignment and perform scheduling by taking advantage of the relaxation in execution time of tasks.

## 1 Introduction

Low power design has been an extremely important issue for embedded systems due to its significant impact on battery life, system density, cooling cost and system reliability. With the shrinking size of the transistors and reducing threshold voltages, the leakage power constitutes an increasing fraction of the total power consumption in modern embedded systems [2]. Thus, power becomes an important constraint in the design specifications of these systems thereby, leading to a significant research in power estimation and low power design.

Embedded computing systems are characterized by the presence of an application specific software running on the specialized processors. The selection of the hardware components for their designing is strongly driven by the power analysis of the system. System models are used to abstract some characteristics from all aspects of the design. The higher the level of abstraction, the greater is the power savings that can be achieved. An abstract system model that contains some functionality information, but not the executable specifications, is the task graph. The system level description consists of both an embedded hardware as well as software. There is a need for power estimation at both the levels.

Our paper focuses on power estimation for embedded software. Accurate power estimation tools are available only for the lower levels of the design - at the circuit level and to a certain extent at the gate level. For an embedded processor, simulation at these levels is slow and it is impractical to evaluate the power consumption of the software. To model the energy consumption of a complex system, it is intuitive to consider individual instructions. As each instruction involves specific processing across various units of the processor, it can result in circuit activity that is the characteristic of each instruction and varies with the instructions. Thus, there is a need to design an approach that takes these features into account.

In this paper, we propose an analytical methodology for power estimation using a graph based analytical approach called HCFG approach [7]. The inputs are the Probability Mass Function distributions of energy for the basic functional units and not the fixed values thereby covering minimum, maximum as well as the average values. We also discuss a HCFG based approach to achieve low power schedule for embedded software for real time systems. The paper is organized as follows. Section 2 presents a review of the relevant work in the literature which are similar to our approach for power analysis and optimization. Section 3 discusses the Hierarchical Concurrent Flow Graph (HCFG) model [7]. In Section 4, we describe an analytical approach for energy estimation. In Section 5, we explore the implementation of our approach on energy aware task scheduling methodologies. Section 6 concludes the paper with discussions on future directions.

## 2 Related Work

Before going into details of our approach, we briefly describe the earlier work done in the area of energy estimation of embedded software. The prospect of combining architecture design and software arrangement at the instruction level has been worked upon to help in the power estimation as reported in [1, 9, 11]. Power consumption in a system is estimated using highly accurate power estimates for the basic modules of the system. Each of the basic modules is modelled for its power consumption. It is a well known fact that with the variations in programs the power consumption of a task graph also varies. The term program here refers to any sequence of code, and does not have to include a logical beginning and an end. The run time of a program may vary according to different input data and initial machine state. There is significant lack of models and tools to analyze this variation.

Authors in [11] estimated the energy consumption by executing an instruction a particular number of times and using the current measurement for the processor during the execution. The inter-instruction effects were estimated by repeated execution of pair of instructions. Using the formulation for power evaluation, $P = I.V_{cc}$ with $P$ denoting the average power, $V_{cc}$ and $I$ the supply voltage and average current respectively; the average power consumed by a processor corresponding to the particular instruction can be estimated. With power value defining the energy consumption rate, the energy consumed by an instruction is given by $E = P \times N \times \tau$, where $N$ represents the clock cycles taken for a sequence of instructions to be executed with $\tau$ as the clock period. Thus, the power consumption for a set of instructions is evaluated by summing up the costs of each instruction along with the inter-instruction effects. This approach gave the desired results but proves to be inefficient, as it requires a large maintenance of data and is valid only for the processor for which it has been measured.

To model the energy consumption of a complex system, it is intuitive to consider individual instructions. Thus, there is a need for a robust and an exhaustive module for the application parameters extraction. Application parameters include ALU operations such as ADD, SUB, MUL, logical operations, load and store operations. The methodology must therefore serve the basic purpose of appropriate identification and extraction of the key parameters to capture the characteristics of the application properly. This information forms the basis of our approach for energy estimation reported in this paper.

Power constraints have become a critical component of the design specifications of the embedded systems that are being used in all walks of life today. Techniques for energy minimization adopted at higher design levels have proven to be more effective than the techniques implemented at lower levels. The power optimization techniques try to provide a solution to the design problem : *Given a task graph and an architecture template for system implementation with several functional units, obtain a mapping of tasks to functional units that minimize energy while maintaining the design constraints*. Power saving techniques at system level include Voltage Selection, that involves selecting an appropriate supply voltage for the processor while meeting desired performance; and Power Management, that involves shutting down of an idle processor.

Voltage selection proves to be a better technique than Power management because the overhead cost involved with the switching of voltages is ignorable if switching does not happen frequently as compared to the cost involved with the switching of the processors. Zhang et al. [8] take real time dependent tasks with deadlines for execution on variable voltage processors. System level implementation has been described as an integration of Task assignment *i.e.* which task runs on which processor, Task execution order, order in which task executes on each processor and Voltage Selection *i.e.* which task is assigned which voltage level. The task assignment and their ordering in first step prepares a ground for voltage selection in the second step. The voltage assignment is based on the fact that higher the voltage level, smaller the execution time but larger the energy consumption for the task execution. The Earliest Deadline First (EDF) scheduling algorithm has been used for scheduling on single processor. Priority based task ordering for multiple processors is being used as EDF does not give optimal solution for multiple processors as tasks will be on multiple paths and affect the paths differently. The priority is defined based on task's deadline, dependencies and usage of processors in the system. Tasks are assigned with the latest finish time so that they and their successors meet the deadlines.

Operating voltage is the deciding factor for the power consumption at the hardware level. So a solution to the power saving problem is to assign the voltage/frequency level for each of the tasks in the given task graph such that the total energy is minimized. This should be achieved without the violation of the timing constraints while assuming that the processors used in the embedded system can exist in one or more operating states, the states being voltage and frequency. By reducing the voltage by a factor of $k$ the energy dissipation can be reduced by a factor of $k^2$ along with the scaling of the frequency of the circuit by a factor of $k$, thereby impacting the performance of the circuit. Thus, the total energy consumption can be optimized within the task execution time constraints by assigning voltage levels to the tasks judiciously. Using the instruction-level energy, $E_{inst}$ and delay, $D_{inst}$ information while taking the task to be a stream of assembly language instructions, the task level energy, $E_{task}$ and delay, $D_{task}$ information can be estimated as described in Algorithm 1.

**Algorithm 1:** Energy calculations
**Input**: number of tasks (count), $E_{inst}$ and $D_{inst}$

**Output**: Task energy and delay
Initialize $E_{task}=0$, $D_{task}=0$;
for $i=1$ to count
$$E_{task} = E_{task} + E_{inst} * (v_{scale})^2 * (f_{scale})$$
$$D_{task} = D_{task} + D_{inst} * (1/v_{scale}) * (f_{scale})$$
end for

Qiu et. al. [3, 10] discuss the voltage assignment problem with guaranteed probability for real time systems. The embedded systems having tasks containing conditional instructions that may have different execution times for different inputs have been explored. The execution time of each node has been modeled as a random variable assuming the Gaussian probability distribution and for the probability values, the Cumulative Distribution Function (CDF) has been sampled. The Voltage assignment with probability (VAP) problem has been defined for selecting an appropriate voltage for each node in the pre-scheduled graph such that the total energy consumption $E$ is minimized while satisfying the timing constraint $L$ with confidence probability $P$.

Dealing with embedded system applications exhibiting large instruction level parallelism (ILP) requires Very Long Instruction Word (VLIW) processors each of which has a certain number of functional units. This design when optimized for peak performance may result in under utilization of the functional units due to variations in ILP. To overcome this, a scheduling algorithm in context of VLIW and clustered VLIW architectures has been proposed by Nagpal and Srikant [5]. The algorithm makes use of the available slack in scheduling instructions such that the idle functional units remain idle for a longer duration while keeping the active units functional. This reduces the number of transitions and increases the idle periods duration, thereby minimizing the leakage energy.

## 3 HCFG: An analytical approach for energy and task-time estimation

Hierarchical Concurrent Flow graph (HCFG) [7] is a technique which supports analysis of flow graphs having hierarchy, concurrency and stochastic nature of the task execution time. In this paper, HCFG [7] approach has been used for the modeling and analysis of the task and task graphs for a given application embedded software. Hierarchy simplifies the description of processes (task graphs) for analysis since it enables many correlated simple tasks to be represented by a single task at higher levels of abstraction. Concurrency allows trade-off between speed and cost as per the availability of resources. Stochastic nature of task parameters like execution time/power generalizes the model and extends its applicability to probabilistic activities. DFLOW is the textual script for describing all the features of HCFG model. The flow or task graph is captured in the form of directed graph.

### 3.1 Time computation

Consider three nodes A, B and C. Their associated edge transmittance are $T_A = p_A.z^{T_A}$, $T_B = p_B.z^{T_B}$ and $T_C = p_C.z^{T_C}$ [7]. For AND concurrency, all sub-tasks must be completed before the next step is performed. The expected completion time is thus given by,
$$E[T_P] = E[Max\{T_A, T_B, T_C\}] \qquad (1)$$

Similarly, for OR-concurrency where the alternate techniques are available for the same problem shifting to the next execution step is possible when either of the techniques has finished executing. The expected completion time in this case is given by,

$$E[T_P] = E[Min\{T_A, T_B, T_C\}]. \tag{2}$$

## 3.2 Power computation

Let $P_A$, $P_B$ and $P_C$ be power estimates of tasks A, B and C respectively and $p_A$, $p_B$ and $p_C$ denote the corresponding probabilities. Their associated edge transmittance are $T_A = p_A.z^{P_A}$, $T_B = p_B.z^{P_B}$ and $T_C = p_C.z^{P_C}$ [7]. The composite transmittance is given by

$$T = T_A + T_B + T_C \tag{3}$$

$$= p_A.z^{P_A} + p_B.z^{P_B} + p_C.z^{P_C} \tag{4}$$

From the above equation, the expected value of power consumption $E[P]$ is evaluated as $\frac{dT}{dz}$ at z=1. Thus,

$$E[P] = p_A * P_A + p_B * P_B + p_C * P_C \tag{5}$$

Consider two scenarios:

1. When A, B, C are lying along alternate paths (OR Concurrency) with $p_A + p_B + p_C = 1$; Equation 5 gives the expected power value for the same.

2. When A, B, C are all concurrent tasks (AND Concurrency); *i.e.*, $p_A = p_B = p_C = 1$. Thus, for concurrent tasks, Equation 5 reduces to:

$$E[P] = P_A + P_B + P_C \tag{6}$$

The probability values are calculated using the profiling information obtained using Trimaran. For each instruction we have the details of the number of times it is being executed. In case of branching, the probabilities of execution of each branch $p_{branch}$ can be obtained as

$$p_{branch} = \frac{number\ of\ times\ that\ branch\ is\ executed}{total\ number\ of\ calls\ to\ the\ instruction} \tag{7}$$

## 4 Power estimation using HCFG

Large ILP available with VLIW processors is facilitated by a certain number of functional units of datapath. Present day embedded software applications posses parallelism across tasks; and hence VLIW architecture are well suited for such applications. Block by block scheduling of the code and binding of the instructions to the available functional units needs to be done by the compiler. We have evaluated the source code using Trimaran suite [6]. Trimaran has the compilation techniques for ILP architectures mainly focusing VLIW architectures. The compiler analyzes the whole program regions and has the capability to perform the mapping between the operations and the corresponding functional units. Sequential steps in our approach used for power estimation have been shown in flow graph of the Figure 1.

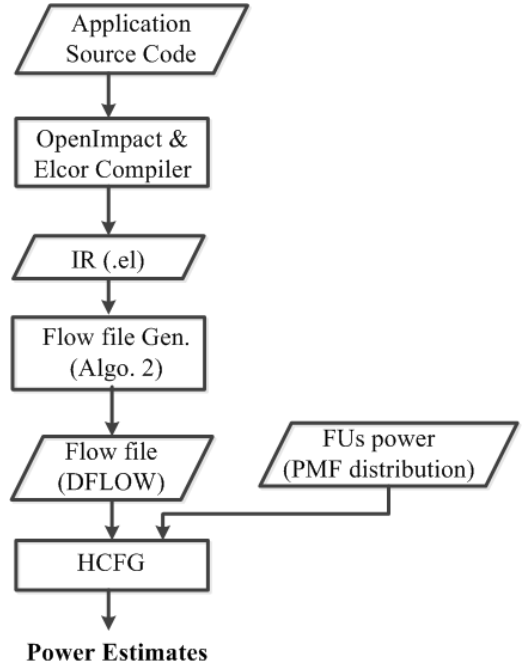

**Figure 1:** Overall methodology for Power Estimation

## 4.1 Algorithm

The approach uses the power estimates of the functional units obtained using PrimePower [4]. The energy model used in Prime Power is the same as the analytical energy model described in paper [2]. Different energy values are obtained using different sets of inputs and from the data thus obtained, distribution for each elementary operation is modeled as certain probability mass function.

The intermediate level application code generated using Trimaran has been used for extracting the flow graph which is to be fed to HCFG as input. This intermediate file provides the instruction level information to predict the usage of functional units and their schedule time. The scheduling of functional units has been performed by the compiler assuming single active functional unit of each type and has been done at the block level. Algorithm 2 is used to obtain the flow files to be fed to the HCFG tool. For each block, the operations being carried out at different time steps are taken in as a sequential flow while, the concurrent operations are defined as either *AND* or *OR* subflows. The operations in a subflow are defined using hierarchical description. Discussing an example case, Consider three concurrent nodes $A$, $B$ and $C$; each having execution time $t$, the transition probability values to be used for the power computation of these task nodes in the flow description file can be calculated as

$$E \times t = E_A \times t + E_B \times t + E_C \times t \tag{8}$$

$$E = E_A + E_B + E_C \tag{9}$$

Thus, the transition probability values for each node is 1. For the case of sequential nodes, the total time becomes $3t$. The equation in this case transforms to

$$E \times 3t = E_A \times t + E_B \times t + E_C \times t \tag{10}$$

$$E = (E_A + E_B + E_C)/3 \tag{11}$$

For a generalized case of $n$ nodes, the expression becomes
$$E = (E_A + E_B + E_C + \cdots + \cdots + E_n)/n \tag{12}$$
Thus, the transition probability values for each node is taken to be $1/n$ for the case of $n$ sequential nodes and '1' for concurrent nodes.

**Algorithm 2**: CDFG Extraction
**Input**: REBEL format file (generated by Elcor compiler)
**Output**: A .flw file (flow-description file)
1. Count the number of each of the blocks($b_k$) in the intermediate file.
2. for $i = 1$ to $b_k$
   Extract the type of functional unit
   The schedule time of each unit
   $max_k$, maximum time steps in each block
   end for
   for $i = 1$ to $b_k$ do
     for $j = 1$ to $max_k$ do
       Group the units having same schedule time
       Output the flow file.
     endfor
   endfor

## 4.2 Experimental setup and Simulation results

The simulations have been performed on a workstation having Core-2 duo Pentium with 2 GB memory, running at 2.8 GHz with Red Hat Linux ES version 4.0 as Operating System. We use benchmark codes of varying sizes and applications on which to demonstrate our methodology for power estimation. The chosen set of benchmarks on which algorithms were run include SPEcint, Mediabench, Netbench, Mibench and other benchmarks. Using Algorithm 4.1 the flow files for input to HCFG tool for power estimation are generated. Table 1 summarizes the energy values for some of the benchmarks.

The *standard deviation* values signify the amount of variations or spread around the mean values. The larger the standard deviation, greater will be the probability of error in power prediction based on mean value. The results tabulated in Table 1 reveal that the average values should not be used but the PMF distributions should be explored for power estimation as the standard deviation values obtained are large. Thus, our approach provides more accurate power analysis than the methodologies proposed in [11, 12] thereby, proving the utility of our approach. The *average power* value can be obtained from the PMF plots using the statistical mean formula

$$P_{avg} = \frac{\sum p_i \times x_i}{\sum x_i} \tag{13}$$

where $p_i$ are the probability values corresponding to the energy values $x_i$. The *most probable value* of the power dissipation for a benchmark will be the value of $x_i$, i.e. the energy value for which $p_i$ is maximum.

**Table 1:** Estimated power for some benchmarks ( $\mu_P$ -Average Power, $\sigma_P$ -Std. deviation)

| Benchmark | Code size (# bb) | $\mu_P$ ($\mu$W) | $\sigma_P$ ($\mu$W) |
|---|---|---|---|
| Allocat | 19 | 40.95 | 39.43 |
| Hyper | 10 | 156.25 | 74.74 |
| Fib | 11 | 12.55 | 27.52 |
| Strcpy | 9 | 34.37 | 23.32 |
| Sha | 49 | 240.02 | 107.56 |
| Qsort | 23 | 22.44 | 38.94 |
| Rawdaudio | 38 | 100.98 | 10.87 |
| Rawcaudio | 45 | 100.96 | 10.78 |

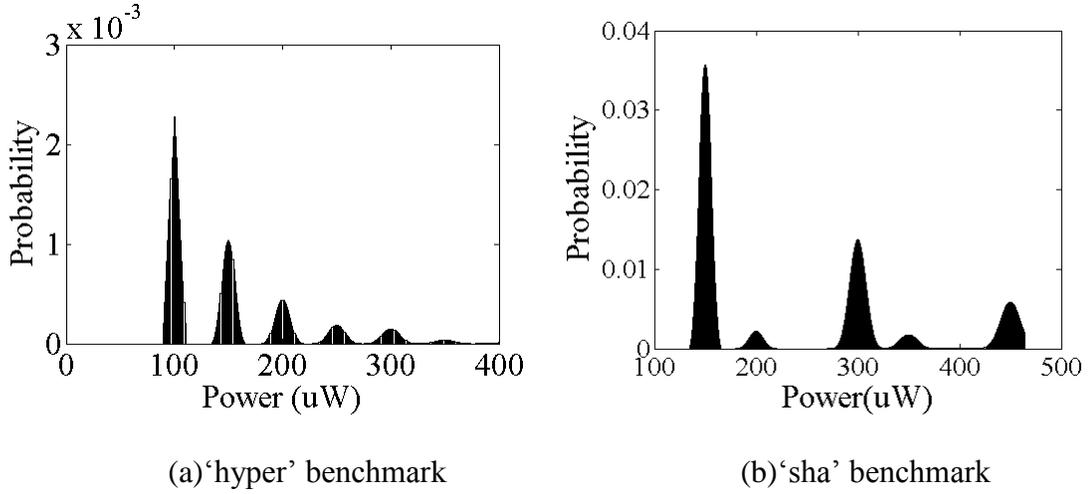

(a) 'hyper' benchmark  (b) 'sha' benchmark

**Figure 2:** PMF plots for some benchmarks

The Probability Mass Function (PMF) plots for 'hyper' and 'sha' benchmarks, are shown in the Figure 2. The energy values are shown on x-axis with their corresponding probabilities on the y-axis. As observed from the Figure 2(a), most of the power values are concentrated between 90-160 $\mu$W, signifying that most probably the power dissipation for 'hyper' benchmark would lie in this range. The most probable energy value for this case is 100 $\mu$W. While considering the PMF plot for power in case of 'sha' benchmark [Figure 2(b)], it can be observed that the power values lie between 120-170 $\mu$W, 150 $\mu$W being the most probable value. The smaller range signifies higher accuracy in power prediction using mean value.

## 5  Energy Aware Task Scheduling

In this section, we present an analytical approach for power optimization of embedded software task graph. The methodology used in this paper exploits the instruction level information extracted using Trimaran to predict the usage of functional units and their schedule time.

## 5.1 Voltage selection based scheduling

We have considered real time dependent tasks with deadlines for execution on variable voltage processors assuming the processor's operatibility at two voltage levels. The higher the voltage level, the faster the execution time and more is the expected energy consumed. The tasks are assigned a latest finish time such that they and their successors meet the deadlines. The precedence constraints for various tasks of the task graph are based on their linking order during compilation. The voltage assignment problem is an optimization problem having large but finite number of solutions. Given $n$ tasks in the task graph and assuming two permitted voltage levels, the total solution space consisting of $2^n$ assignments has been explored. The Algorithm 3 returns the best and worst case voltage assignments possible within the deadline.

**Algorithm 3:** Voltage selection based scheduling
**Input**: $n$ voltage levels, Task graph, timing constraint
**Output**: An optimal voltage assignment
1. Explore all the possible voltage assignments for each basic block;
2. Obtain the flow files (HCFG .flw file) for each possible schedule;
3. Select using results from HCFG, those schedules getting completed within the deadline;
4. Obtain the best and worst case voltage assignments comparing their power PMFs.

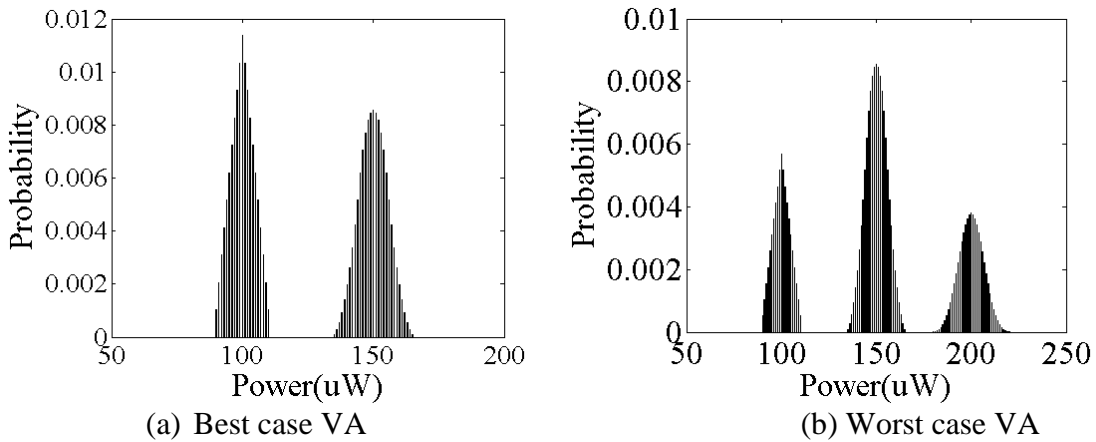

(a) Best case VA  (b) Worst case VA

**Figure 3:** PMF plot for estimated power for 'hyper' benchmark

The concept behind the best case voltage assignment (VA) is, the maximum time slack utilization. In this case, maximum number of tasks are assigned the lower voltage and thus the energy consumption is minimized. While in case of worst case voltage assignment, voltages are assigned to the tasks such that the task completion time is minimum. The energy consumption is maximum in this case. This is truly reflected in PMF plots shown in the Figure 3 for the best and worst case voltage assignments for 'hyper' benchmark. An approximate estimation of the energy savings can be done by calculating the difference between the average power values obtained for these voltage assignments. The theoretical values of energy savings have been calculated [8] using,

$$Energy\ Savings = SN \times (V_h^2 - V_l^2) \qquad (14)$$

where,

$SN$ = number of slowed down cycles
$V_h$ = 1.8 V, high level voltage
$V_l$ = 0.9 V, low level voltage.

**Table 2:** Energy savings with voltage selection methodology (SN- Slowdown cycles, MP- Most probable value of energy, $\mu_P$- Average power)

| Benchmark | SN | Best case VA | | Worst case VA | | Energy savings ($\mu$W) | |
|---|---|---|---|---|---|---|---|
| | | MP ($\mu$W) | $\mu_P$ ($\mu$W) | MP ($\mu$W) | $\mu_P$ ($\mu$W) | Estimation approach | Theoretically |
| Fib | 2 | 45 | 17.025 | 50 | 21.83 | 4.805 | 4.80 |
| Dag | 7 | 125 | 125 | 150 | 145 | 20 | 17.01 |
| Alloca | 7 | 55 | 70.56 | 90 | 91.23 | 20.67 | 17.01 |
| Hyper | 18 | 100 | 100 | 150 | 149.99 | 49.9 | 43.2 |
| Strcpy | 9 | 55 | 56.41 | 75 | 78.70 | 22.3 | 21.87 |

The number of slowed down cycles has been calculated using the best and worst case voltage assignments obtained for each task. Table 2 reveals that the variations in energy savings from the theoretically calculated results is small for the benchmarks where the PMF has a smaller range. But a significant difference is achieved for the task graphs where the PMF distributions for individual tasks have a larger variation. This shows that our approach gives more accurate analysis than the analysis based on mean values.

## 5.2 Time constrained multiprocessor scheduling

The objective here is to maximize the utilization of the available time slack for each task. We aim to activate minimum number of the available processors so that the leakage energy is minimized. This methodology in a way minimizes the expected total energy consumption while satisfying the timing constraint with a guaranteed confidence probability. For soft real time analysis, the deadlines of each task can be relaxed to the extent that the complete task graph satisfies the desired confidence probability.

**Algorithm 4:** Time constrained multiprocessor scheduling for real time systems
**Input**: $n$ processors, task graph, confidence probability
**Output**: Optimal schedule
1. Schedule the task graph, starting with the minimum number of processors;
2. Select the schedules completing within the deadline satisfying the confidence probability;
3. For each processor obtain the voltage assignment using Algorithm 3;
4. Output the power PMF for each processor's schedule.

Considering the multi-processor system, the minimum number of processors required to schedule a task set is given by $\lceil \frac{T_t}{D} \rceil$, where $T_t$, is the total computation time of the tasks in the given task graph and $D$, the deadline. The energy savings show a steep rise when the number of processors is small, but with increase in the number of processors energy savings does not change significantly, because of limited parallelism among the tasks. The leakage energy optimal schedule has been obtained for a given complete task graph using Algorithm 4. Table 3 summarizes the results for some benchmarks. Columns show the number of tasks, number of task cycles, the deadline for each processor and the number of resources required for scheduling under such constraints. The execution time of each of the processors has also been listed. This is the maximum time for which each processor is active. The maximum time limit under which the task graph will be scheduled has been evaluated for each application using HCFG.

**Table 3:** Completion time for multiprocessor-time-constrained-scheduling

| Benchmark | # Tasks | # Cycles | Deadline | # Resources | Execution time $P_1$ | $P_2$ | $P_3$ | Overall time Est.Approach |
|---|---|---|---|---|---|---|---|---|
| Epic | 10 | 1380 | 966 | 2 | 59 | 421 | | 1042 |
| Pegwitenc | 10 | 2530 | 1265 | 3 | 650 | 920 | 960 | 1344 |
| Mpeg2enc | 18 | 3933 | 2359 | 3 | 1025 | 1905 | 1003 | 2386 |
| Decode | 6 | 750 | 450 | 2 | 375 | 375 | | 572 |
| Basicmath | 4 | 1332 | 799 | 2 | 667 | 665 | | 657 |

Figure 5 shows the PMF plots of estimated power for the best case voltage assignment for the 'epic' benchmark application scheduled on two processors, $P_1$ and $P_2$ respectively. The average power value for this schedule will be the sum of the average power values for processors $P_1$ and $P_2$. Similarly, the power distribution on each processor for different benchmarks can be obtained and thus, the resultant average power for a particular application can be estimated.

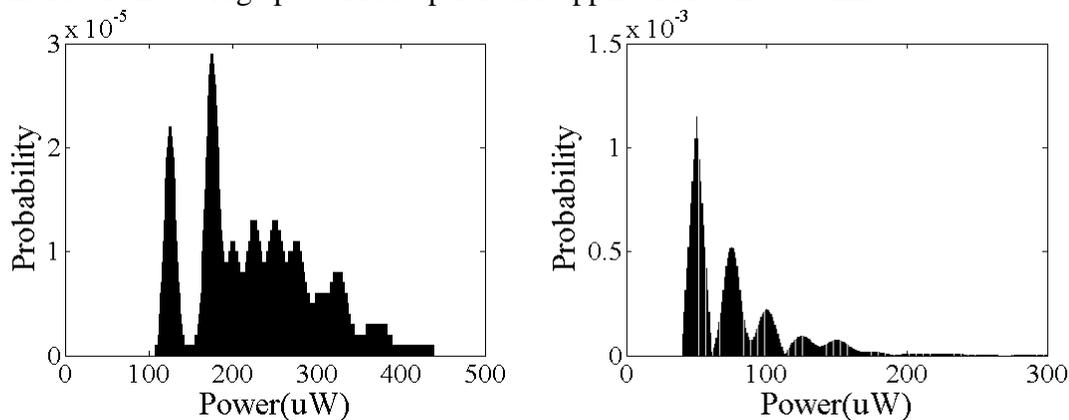

**Figure 4:** PMF plot for estimated power for 'epic' benchmark on $P_1$ and $P_2$

## 6 Conclusions and Future Work

The motivation behind comprehensive power analysis is that it provides insight into energy consumption pattern in processors. It helps in verifying if an embedded design meets its power constraints. This can also be used to guide the design of embedded software such that it meets the constraints. We have presented a novel approach for power estimation of embedded software using the Control Data Flow Graph (CDFG) for task and task graph. Trimaran compiler has been used for the extraction of application parameters whereas PrimePower has been used to obtain power estimates for elementary functional units. We have also presented an analytical approach to obtain an energy optimal voltage assignment for a task graph. The results prove its effectiveness for the cases where the probability-energy distribution curve has a larger range. We have also proposed a multiple-tasks scheduling technique that exploits data parallelism of tasks targeted for scalable multiprocessors. The objective is to find a schedule that respects all the constraints e.g. precedence, communication, deadline etc. By taking advantage of the allowed relaxation in execution time of tasks, an energy optimal voltage assignment and scheduling has been achieved. Future work along these lines may include designing of a more realistic processor model that takes into account the effects of cache memory. Also, the algorithms that manage energy slack may be developed for real time systems. Genetic Algorithm can be used to find an optimal voltage assignment for a task graph in a multi-processor system.

# References

[1] V. S. Lapinskii, D. Veciana, and G. A., "Cluster assignment for high-performance embedded VLIW processors," in *ACM Transactions on Design Automation of Electronic Systems*, 2002, pp. 430–454.

[2] S. Dropsho, V. Kursun, D. H. Albonesi, S. Dwarkadas, and E. G. Friedman, "Managing static leakage energy in microprocessor functional units," in *Proceedings of the 35th annual ACM/IEEE international symposium on Microarchitecture*. 1em plus 0.5em minus 0.4em IEEE Computer Society Press, 2002, pp. 321–332.

[3] M. Qiu, J. Wu, F. Hu, S. Liu, and L. Wang, "Voltage Assignment for Soft Real-Time Embedded Systems with Continuous Probability Distribution," in *RTCSA*, 2009, pp. 413–418.

[4] Synopsys, "PrimePower tutorial," http://www.synopsys.com.

[5] R. Nagpal and Y. N. Srikant, "Compiler-assisted leakage energy optimization for clustered VLIW architectures," in *Proceedings of the 6th ACM & IEEE International conference on Embedded software*. 1em plus 0.5em minus 0.4em ACM, 2006, pp. 233–241.

[6] "Trimaran Compiler," http://www.trimaran.org.

[7] V. Sahula and C. P. Ravikumar, "The Hierarchical Concurrent Flow Graph Approach for Modeling and Analysis of Design Processes," in *VLSI Design*, 2001, pp. 91–96.

[8] Y. Zhang, X. S. Hu, and D. Z. Chen, "Task scheduling and voltage selection for energy minimization," in *Proceedings of the 39th annual Design Automation Conference*. 1em plus 0.5em minus 0.4em ACM, 2002, pp. 183–188.

[9] Y.-T. S. Li and S. Malik, "Performance Analysis of Embedded Software Using Implicit Path Enumeration," in *Design Automation Conference*, 1995, pp. 456–461.

[10] M. Qiu, Z. Jia, C. Xue, Z. Shao, and E. H.-M. Sha, "Voltage Assignment with Guaranteed Probability Satisfying Timing Constraint for Real-time Multiproceesor DSP," *VLSI Signal Processing*, vol. 46, no. 1, pp. 55–73, 2007.

[11] V. Tiwari, S. Malik, and A. Wolfe, "Power analysis of embedded software: a first step towards software power minimization," in *ICCAD*, 1994, pp. 384–390.

[12] M. T.-C. Lee, V. Tiwari, S. Malik, and M. Fujita, "Power analysis and minimization techniques for embedded DSP software," *IEEE Transactions on VLSI Systems*, vol. 5, no. 1, pp. 123–135, 1997.